\documentclass[preprint,showpacs,preprintnumbers,amsmath,amssymb]{revtex4}

\usepackage{graphicx}
\usepackage{dcolumn}
\usepackage{bm}
\begin{document}
%
\title{Alternative pathways of dewetting for a thin liquid two-layer film}

\author{Andrey Pototsky}
\affiliation{Lehrstuhl f{\"u}r Theoretische Physik II,
 Brandenburgische Technische Universit\"at Cottbus,
 Erich-Weinert-Stra{\ss}e 1, D-03046 Cottbus, Germany}

\author{Michael Bestehorn}
\affiliation{Lehrstuhl f{\"u}r Theoretische Physik II,
 Brandenburgische Technische Universit\"at Cottbus,
 Erich-Weinert-Stra{\ss}e 1, D-03046 Cottbus, Germany}

\author{Uwe Thiele}
\affiliation{Max-Planck-Institut f\"ur Physik komplexer Systeme,
 N\"othnitzer Stra{\ss}e 38, D-01187 Dresden, Germany}

\author{Domnic Merkt}
\affiliation{Lehrstuhl f{\"u}r Theoretische Physik II,
 Brandenburgische Technische Universit\"at Cottbus,
 Erich-Weinert-Stra{\ss}e 1, D-03046 Cottbus, Germany}

\begin{abstract}
We consider two stacked 
ultra-thin layers of different liquids on a solid substrate.
Using long-wave theory, we derive coupled evolution equations for the free 
liquid-liquid and liquid-gas interfaces. Depending on the 
long-range van-der-Waals forces
and the ratio of the layer thicknesses, the system follows different pathways
of dewetting. The instability may be driven by
varicose or zigzag modes and leads to film rupture either at the 
liquid-gas interface or at the substrate. 
We predict that
the faster layer drives the evolution and may accelerate the
rupture of the slower layer by orders of magnitude promoting thereby the
rupture of rather thick films.
\end{abstract}


\maketitle
Instability phenomena in ultra-thin soft matter films with
thicknesses below $100$\,nm became relevant
mainly because they obstruct the fabrication
of homogeneous coatings \cite{Reit92}. The interest was further boosted
by the possibility to control such processes and,
to use them to manu\-facture functional layers on 
the nanometer scale \cite{Mert97a,TMP98}. 
The stability of  ultra-thin films is dominated by the
effective molecular interactions between
the substrate and the film surface \cite{RuJa73}. 
They represent, for instance, long-range van-der-Waals
forces which increase (decrease) the pressure in the film if
they are attractive (repulsive) \cite{Isra92}. However, to determine 
the emerging length scale and pattern for unstable films a study of 
the film dynamics is required. 
Using a film thickness evolution equation obtained by long-wave approximation
\cite{ODB97}, the dewetting of a single layer of liquid is now reasonably 
well understood (see e.g.\ Ref.\,\cite{Thie03}). 

However, little is known on the behaviour
of two stacked ultra-thin layers of simple or polymeric liquid on a solid substrate
(see Fig.\,\ref{fig1}). 
Such a two-layer film allows for richer dynamics
than an one-layer system because, both, the free liquid-liquid and the free 
liquid-gas interface evolve in a coupled way.  The evolution is driven
by the effective molecular interactions between {\it all} 
the three interfaces separating
the four material layers: substrate, liquid$_1$, liquid$_2$ and ambient gas.
Although experimental studies investigated different aspects of dewetting
for two-layer films like interface instabilities or the growth of holes 
\cite{FCW95,LPHK96,PWHC97,Sfer97,Sfer98,DRSS98,RMSH00,MSS03}
up to now no general theoretical description of the interface dynamics has
been given \cite{note2}.  
The case of small interface deflections was investigated in Ref.\,\cite{BMR93} 
for a thickness of the lower layer, $d_1$, much larger than
that of the upper one $(d_2-d_1)$. 

The most intricate questions for the first stage of dewetting of a two-layer
system are {\it which} interface will become unstable, 
{\it where} does the film rupture, and {\it how long} will it take.
This will determine the observability of the instability and the final morphology 
of the film. Experiments found roughening of the
liquid-liquid interface \cite{Sfer98} or an instability of the liquid-gas
interface \cite{FCW95,MSS03}. Holes that evolve solely in the upper 
layer were also studied \cite{LPHK96,PWHC97}.

In this Rapid Communication, we derive and analyse coupled long-wave
evolution equations for the two interfaces that are
valid for all interface deflections and thickness ratios.
We show that solely by changing the thickness ratio of the layers
one switches between different dominant instability 
modes. This leads to drastic changes of
the pathway of dewetting from rupture
at the substrate to rupture at the liquid-liquid interface 
(see Fig.\,\ref{fig4} below). Remarkably, for systems composed of two layers
of very different thickness, i.e.\ with very different time scales
for the rupture of the individual layers, 
the faster layer drives the evolution and accelerates the 
growth of surface modulations of the slower layer by orders of magnitude.
We illustrate our results for two-layer systems of polystyrene (PS)
and polymethylmetacrylate (PMMA) with silicon (Si) or 
silicon oxide (SiO) as substrate
like studied experimentally in Refs.\,\cite{LPHK96,Sfer97,MSS03}.

We believe, our model can be used not only for 
the description of two-layer experiments with simple or polymeric 
liquids, but using an appropriate free
energy functional also for a liquid film 
on a substrate with a stable but soft coating like a polymer 
brush \cite{RSAA96} and, including driving terms, for studies of 
the transport of liquid droplets
in liquid-liquid microfluidic systems \cite{VPB03}.

{\it Coupled film thickness equations:}\hspace{.3cm}
We obtain evolution equations for the film thicknesses $h_1$ and $h_2$ 
by simplifying the Navier-Stokes equations employing long-wave approximation 
\cite{ODB97}.  Thereby a no-slip condition at the substrate, 
the continuity of the velocity field 
and the balance of the stress-tensors at the liquid-liquid and 
liquid-gas interfaces are used. 
Considering an isothermal two-layer system where both layer thicknesses 
are smaller than $100$\,nm,
we neglect gravity and solely focus on the effective molecular interaction. 
For simplicity we only regard  non-retarded long-range van-der-Waals forces
resulting from dipole-dipole interactions between apolar materials.
However, inclusion of other forces, like e.g.\ 
short-range polar forces \cite{Shar93}, 
or of slip boundary conditions \cite{ODB97} (that may be necessary
for polymer films) is straightforward as for one-layer films.
Details of the derivation will be presented elsewhere.
We obtain \cite{note2}
\begin{eqnarray}
\frac{\partial h_1}{\partial t} \,&=&\, \nabla \left(Q_{11}  
\nabla \frac{\delta F}{\delta h_1}  + 
Q_{12} \nabla \frac{\delta F}{\delta h_2}  \right) \nonumber\\
\frac{\partial h_2}{\partial t} \,&=&\, \nabla \left(Q_{21} 
\nabla \frac{\delta F}{\delta h_1}  + 
Q_{22} \nabla \frac{\delta F}{\delta h_2}  \right),
\label{TWOL_EV}
\end{eqnarray}
where $\delta F/\delta h_i$ with $i=1,2$ denotes functional derivatives
of the total energy of the system
\begin{equation}
F \,=\, \int [ \rho_{\text{s}} +\rho_{\text{VW}}]\,d{\bm x}\,.
\label{energ}
\end{equation}
It contains the densities of the surface energy
$\rho_{\text{s}} = \frac{1}{2}[\sigma_1 
(\nabla h_1)^2 + \sigma_2 (\nabla h_2)^2 ]$,
and of the energy for the van-der-Waals interaction  
$\rho_{\text{VW}} = - A_{g21s}/(12 \pi h_2^2) - A_{21s}/(12 \pi h_1^2) - 
A_{12g}/[12 \pi (h_2-h_1)^2]$. The surface tensions
$\sigma_1$ and $\sigma_2$ belong to
the liquid-liquid and liquid-gas interface, respectively.
$A_{g21s}$, $A_{21s}$ and $A_{12g}$ are four- and three-index Hamaker 
constants, with subscripts s, 1, 2 and g refering to the substrate,
liquid$_1$, liquid$_2$ and gas, respectively \cite{note1}.
The symmetric matrix of the positive mobility factors $Q_{ik}$ reads
\begin{equation}
{\bm Q} =\frac{1}{3 \mu_1}
\left(
\begin{array}{cc}  
h_1^3& \frac{3 h_1^2}{2} \left(h_2 -\frac{h_1}{3} \right) \\[.3ex]
\frac{3 h_1^2}{2} \left(h_2 -\frac{h_1}{3} \right) &\,
\frac{(h_2 -h_1)^3 (\mu_1-\mu_2)}{\mu_2} +h_2^3
\end{array}
\right)
\label{QQ}
\end{equation}
where $\mu_1$ and $\mu_2$ are the viscosities of liquid$_1$ and 
liquid$_2$, respectively.
Note, that for $d_2-d_1 \ll d_1$ and for small surface deflections 
Eqs.\,(\ref{TWOL_EV}) simplify to those of Ref.\,\cite{BMR93}.
Assuming two identical liquids, Eqs.\,(\ref{TWOL_EV}) reduce to
the well known one-layer equation \cite{ODB97}.

To compare to the well understood one-layer systems 
we non-dimensionalize Eqs.\,(\ref{TWOL_EV}) using scales derived from the
upper layer as an effective one-layer system. We scale $\bm x$ with 
$\lambda_{\text{up}}=4\pi(d_2-d_1)^2 \sqrt{\pi \sigma_2 / |A_{12g}|}$, $h_i$ with $d_2-d_1$
and $t$ with $\tau_{\text{up}}=48\pi^2 \mu_2 \sigma_2(d_2-d_1)^5 / A_{12g}^2$.
The corresponding energy scale is $|A_{12g}|/16\pi^3(d_2-d_1)^2$.
The ratios of the mean thicknesses, 
surface tensions and viscosities are $d = d_2/d_1$, $\sigma = \sigma_2 / \sigma_1$
and $\mu=\mu_2/\mu_1$, respectively. To compare with the lower layer as 
effective one-layer system one introduces in an analogous way the length scale
$\lambda_{\text{low}}$ and time scale $\tau_{\text{low}}$.

We simulate the coupled time evolution of $h_1$ and $h_2$, 
Eqs.\,(\ref{TWOL_EV}), in an one-dimensional 
domain using a semi-implicit time integration scheme and periodic
boundary conditions. Initial conditions consist of flat layers with
an imposed noise of amplitude 0.001. 
Alternative pathways of dewetting that occur for different thickness ratios $d$
are presented in Fig.\,\ref{fig4} using a Si/PMMA/PS/air system as an example. 
Fig.\,\ref{fig4}\,(a) shows that for a relatively small $d=1.4$
the two interfaces start to evolve deflections that are in anti-phase
indicating the dominance of a varicose mode.
When the liquid-gas interface approaches the liquid-liquid interface
the latter starts to move downwards due to dynamical effects. This pathway
leads to rupture of the upper layer, i.e.\ at the liquid-gas interface.
On the contrary, Fig.\,\ref{fig4}\,(b) shows that for a larger
$d=2.4$ the growing deflections of the two interfaces are in phase
indicating the dominance of a zigzag mode.
As a consequence, here the lower layer ruptures, i.e. rupture occurs 
at the substrate. 

Note, that in both, Fig.\,\ref{fig4}\,(a) and (b), at the moment of rupture
the respective non-ruptured layer is also in an advanced stage of its evolution
leading to subsequent rupture. This is remarkable because their 
time scales as effective one-layer systems are 15 times (Fig.\,\ref{fig4}\,(a))
and 35 times (Fig.\,\ref{fig4}\,(b)) slower than the time scales for the 
respective fast layer.
The ratio of the time scales $\tau_{\text{up}}/\tau_{\text{low}}$ 
is proportional to $(d-1)^5$, i.e.\
for a lower layer ten times thicker than the upper one the rupture time
of the lower layer is about five orders of magnitude larger than the
one of the upper layer. However, a simulation for a Si/PMMA/PS/air system
with $d_1=10$ and $d_2-d_1=1$ shows that at rupture of the upper layer at
$t=0.61\tau_{\text{up}}=3.99\times10^{-5}\tau_{\text{low}}$ 
the lower layer already evolved a depression of
one fourth of its thickness. If the lower layer is the fast one, the 
effect also exists but is less pronounced.

In both cases, the acceleration of the rupture of the slower 
layer is caused by the direct 
coupling of the layers via the liquid-liquid interface.
The fast evolution of the thinner layer deforms the interface and brings
the thicker layer beyond the slow linear stage of its evolution.
If the upper layer is the driving layer the process is in addition dynamically 
enforced because the liquid-liquid interface is 'pushed away' by the 
advancing liquid-gas interface. 

{\it Linear stability analysis:}\hspace{.3cm} 
Deeper understanding of the different pathways can be reached by studying 
the linear stability of the initial flat layers.
We linearize Eqs.\,(\ref{TWOL_EV}) for small disturbances
$\chi_i \exp{(\beta t)} \cos(kx)$ for $i=1,2$,
where $k$, $\beta$ and ${\bm \chi}=(\chi_1,\chi_2)$ 
are the wave number, growth rate and amplitudes of the disturbance, 
respectively. 
The dispersion relation $\beta(k)$ is obtained solving the resulting
eigenvalue problem 
$(k^2 {\bm Q}\cdot {\bm E}(k)+\beta {\bm I}) {\bm \chi} = 0$, where
${\bm Q}$ is the scaled mobility matrix and
${\bm E}$ is derived from the free energy Eq.\,(\ref{energ}) 
as $E_{ij}= \partial_{h_i h_j} \rho_{\text VW} + \delta_{ij}\tilde{\sigma}_i k^2$
($\tilde{\sigma}_1=1$, $\tilde{\sigma}_2=\sigma$, 
and $\delta_{ij}=1$ for $i=j$ and zero otherwise).
The stability threshold shown as solid line in Fig.\,\ref{fig2}
is determined by ${\bm E}$ for disturbances of 
infinite wavelength, i.e.\ $k=0$.
The system is linearly stable for
\begin{eqnarray}
\det {\bm E} > 0 \quad\text{and}\quad E_{11}> 0 
\quad\text{at}\quad k=0.
\label{STAB}
\end{eqnarray}
An instability sets in if at least one of the conditions (\ref{STAB}) is violated.
This implies, that the two-layer film can be unstable 
even if both, $\partial_{h_1 h_1}\rho_{\text VW}$ and
$\partial_{h_2 h_2}\rho_{\text VW}$ are positive, i.e.\ if the effective 
one-layer systems related to these terms are both stable. 

Fixing the Hamaker constants, i.e.\ the combination of materials,
and changing $d$ one finds a line (trajectory) in the stability
diagram Fig.\,\ref{fig2} as shown for a variety of experimentally studied
systems. Interestingly, for van-der-Waals interactions calculated as detailed 
in \cite{note1} one can show that 
such a trajectory {\it can not} cross the stability threshold,
i.e.\ it is not possible to stabilize such a two-layer system by solely
changing $d$.
For instance, for the Si/PMMA/PS/air system 
the second condition in (\ref{STAB}) is violated for all $d$ and the system is 
always unstable. At $d = 1$, i.e.\ for a vanishing upper layer,
the system is on the boundary between the one- and the two-mode regions
(dashed line).
For $1 < d < 2.3$ the unstable mode is an asymmetric varicose mode.
A corresponding dispersion relation  $\beta(k)$ 
is shown for $d=1.4$ in Fig.\,\ref{fig3}
(cp.\ the time evolution in Fig.\,\ref{fig4}\,(a)).
For $d > 2.3$, i.e.\ for smaller thicknesses of the lower layer, 
the unstable mode is an asymmetric zigzag mode. Fig.\,\ref{fig3} gives
$\beta(k)$ for
$d=2.4$ corresponding to the time evolution shown in Fig.\,\ref{fig4}\,(b). 
For the fastest mode the zigzag mode is strongly asymmetric, i.e.\ 
the deflection of the liquid-liquid
interface dominates the linear stage of the evolution. 
Note, that $\beta(k)$ and the dominant mode type
depend on $\sigma$ and $\mu$, whereas the stability {\it does not}.

Further on, the simultaneous action of the van-der-Waals forces between the 
three interfaces allows for dispersion relations with two maxima.
An experimental system showing this unusual form of $\beta(k)$ can be realized
with a substrate that is less polarisable than both layers.
This is the case for the SiO/PMMA/PS/air system \cite{note1}.
A dispersion relation showing maxima of equal height
is given for $d=2.16$ and $\sigma=10$ in Fig.\,\ref{fig3}.
The maxima at small and large $k$ correspond to strongly asymmetric 
zigzag and varicose modes, respectively.  
This implies that the 
larger (smaller) wavelength will predominantly be seen
at the liquid-gas (liquid-liquid) interface (Fig.\,\ref{fig5}\,(a)). 
Increasing (decreasing) the ratio of the surface tensions strengthens
the smaller (larger) wavelength. This implies that solely changing
$\sigma$ by adding an otherwise passive surfactant one can switch from an
evolution entirely dominated by the liquid-liquid interface to one
dominated by the liquid-gas interface.
This illustrates Fig.\,\ref{fig5} by single snapshots from the
non-linear time evolutions for different $\sigma$.
 
To conclude, we have derived coupled evolution equations for
a thin liquid two-layer film driven by long-range van-der-Waals 
forces. The system represents the most general form 
of coupled evolution equations for two conserved order parameter fields
in a relaxational situation and is apt to 
describe a broad variety of experimentally studied two-layer systems Refs.\,\cite{ThJB04}.
Linear and non-linear analysis have shown that the mobilities have no influence
on the stability threshold, but determine the length and time scales 
of the dynamics. We have shown that for a two-layer
system {\it both} interface deflection modes - zigzag and varicose mode -
may be unstable and lead to rupture at the substrate or the
liquid-liquid interface. Remarkably, the faster layer accelerates the 
evolution of the slower layer even if the latter is rather thick implying that
its rupture time may be shortened by orders of magnitude.

\newpage

\begin{thebibliography}{20}
\expandafter\ifx\csname natexlab\endcsname\relax\def\natexlab#1{#1}\fi
\expandafter\ifx\csname bibnamefont\endcsname\relax
  \def\bibnamefont#1{#1}\fi
\expandafter\ifx\csname bibfnamefont\endcsname\relax
  \def\bibfnamefont#1{#1}\fi
\expandafter\ifx\csname citenamefont\endcsname\relax
  \def\citenamefont#1{#1}\fi
\expandafter\ifx\csname url\endcsname\relax
  \def\url#1{\texttt{#1}}\fi
\expandafter\ifx\csname urlprefix\endcsname\relax\def\urlprefix{URL }\fi
\providecommand{\bibinfo}[2]{#2}
\providecommand{\eprint}[2][]{\url{#2}}

\bibitem[{\citenamefont{Reiter}(1992)}]{Reit92}
\bibinfo{author}{\bibfnamefont{G.}~\bibnamefont{Reiter}},
  \bibinfo{journal}{Phys. Rev. Lett.} \textbf{\bibinfo{volume}{68}},
  \bibinfo{pages}{75} (\bibinfo{year}{1992}).

\bibitem[{\citenamefont{Mertig et~al.}(1998)\citenamefont{Mertig, Thiele,
  Bradt, Klemm, and Pompe}}]{Mert97a}
\bibinfo{author}{\bibfnamefont{M.}~\bibnamefont{Mertig et~al.}},
  \bibinfo{journal}{Appl. Phys. A} \textbf{\bibinfo{volume}{66}},
  \bibinfo{pages}{S565} (\bibinfo{year}{1998}).

\bibitem[{\citenamefont{Thiele et~al.}(1998)\citenamefont{Thiele, Mertig, and
  Pompe}}]{TMP98}
\bibinfo{author}{\bibfnamefont{U.}~\bibnamefont{Thiele}},
  \bibinfo{author}{\bibfnamefont{M.}~\bibnamefont{Mertig}}, \bibnamefont{and}
  \bibinfo{author}{\bibfnamefont{W.}~\bibnamefont{Pompe}},
  \bibinfo{journal}{Phys. Rev. Lett.} \textbf{\bibinfo{volume}{80}},
  \bibinfo{pages}{2869} (\bibinfo{year}{1998}).

\bibitem[{\citenamefont{Ruckenstein and Jain}(1974)}]{RuJa73}
\bibinfo{author}{\bibfnamefont{E.}~\bibnamefont{Ruckenstein}} \bibnamefont{and}
  \bibinfo{author}{\bibfnamefont{R.}~\bibnamefont{Jain}}, \bibinfo{journal}{J.
  Chem. Soc. Faraday Trans. II} \textbf{\bibinfo{volume}{70}},
  \bibinfo{pages}{132} (\bibinfo{year}{1974}).

\bibitem[{\citenamefont{Israelachvili}(1992)}]{Isra92}
\bibinfo{author}{\bibfnamefont{J.~N.} \bibnamefont{Israelachvili}},
  \emph{\bibinfo{title}{Intermolecular and Surface Forces}}
  (\bibinfo{publisher}{Academic Press}, \bibinfo{address}{London},
  \bibinfo{year}{1992}).

\bibitem[{\citenamefont{Oron et~al.}(1997)\citenamefont{Oron, Davis, and
  Bankoff}}]{ODB97}
\bibinfo{author}{\bibfnamefont{A.}~\bibnamefont{Oron}},
  \bibinfo{author}{\bibfnamefont{S.~H.} \bibnamefont{Davis}}, \bibnamefont{and}
  \bibinfo{author}{\bibfnamefont{S.~G.} \bibnamefont{Bankoff}},
  \bibinfo{journal}{Rev. Mod. Phys.} \textbf{\bibinfo{volume}{69}},
  \bibinfo{pages}{931} (\bibinfo{year}{1997}).

\bibitem[{\citenamefont{Thiele}(2003)}]{Thie03}
\bibinfo{author}{\bibfnamefont{U.}~\bibnamefont{Thiele}},
  \bibinfo{journal}{Eur. Phys. J. E} \textbf{\bibinfo{volume}{12}},
  \bibinfo{pages}{409} (\bibinfo{year}{2003})  \bibinfo{note}{and other
contributions of the Focus Point in \bibinfo{journal}{Eur. Phys. J. E}
\textbf{\bibinfo{volume}{12}}}.

\bibitem[{\citenamefont{Faldi et~al.}(1995)\citenamefont{Faldi, Composto, and
  Winey}}]{FCW95}
\bibinfo{author}{\bibfnamefont{A.}~\bibnamefont{Faldi}},
  \bibinfo{author}{\bibfnamefont{R.~J.} \bibnamefont{Composto}},
  \bibnamefont{and} \bibinfo{author}{\bibfnamefont{K.~I.} \bibnamefont{Winey}},
  \bibinfo{journal}{Langmuir} \textbf{\bibinfo{volume}{11}},
  \bibinfo{pages}{4855} (\bibinfo{year}{1995}).

\bibitem[{\citenamefont{Lambooy et~al.}(1996)\citenamefont{Lambooy, Phelan,
  Haugg, and Krausch}}]{LPHK96}
\bibinfo{author}{\bibfnamefont{P.}~\bibnamefont{Lambooy}},
  \bibinfo{author}{\bibfnamefont{K.~C.} \bibnamefont{Phelan}},
  \bibinfo{author}{\bibfnamefont{O.}~\bibnamefont{Haugg}}, \bibnamefont{and}
  \bibinfo{author}{\bibfnamefont{G.}~\bibnamefont{Krausch}},
  \bibinfo{journal}{Phys. Rev. Lett.} \textbf{\bibinfo{volume}{76}},
  \bibinfo{pages}{1110} (\bibinfo{year}{1996}).

\bibitem[{\citenamefont{Pan et~al.}(1997)\citenamefont{Pan, Winey, Hu, and
  Composto}}]{PWHC97}
\bibinfo{author}{\bibfnamefont{Q.}~\bibnamefont{Pan}},
  \bibinfo{author}{\bibfnamefont{K.~I.} \bibnamefont{Winey}},
  \bibinfo{author}{\bibfnamefont{H.~H.} \bibnamefont{Hu}}, \bibnamefont{and}
  \bibinfo{author}{\bibfnamefont{R.~J.} \bibnamefont{Composto}},
  \bibinfo{journal}{Langmuir} \textbf{\bibinfo{volume}{13}},
  \bibinfo{pages}{1758} (\bibinfo{year}{1997}).

\bibitem[{\citenamefont{Sferrazza et~al.}(1997)\citenamefont{Sferrazza, Xiao,
  Jones, Bucknall, Webster, and Penfold}}]{Sfer97}
\bibinfo{author}{\bibfnamefont{M.}~\bibnamefont{Sferrazza et~al.}},
  \bibinfo{journal}{Phys. Rev. Lett.} \textbf{\bibinfo{volume}{78}},
  \bibinfo{pages}{3693} (\bibinfo{year}{1997}).
%
\bibitem[{\citenamefont{Sferrazza et~al.}(1998)\citenamefont{Sferrazza,
  Heppenstall-Butler, Cubitt, Bucknall, Webster, and Jones}}]{Sfer98}
\bibinfo{author}{\bibfnamefont{M.}~\bibnamefont{Sferrazza et~al.}},
  \bibinfo{journal}{Phys. Rev. Lett.} \textbf{\bibinfo{volume}{81}},
  \bibinfo{pages}{5173} (\bibinfo{year}{1998}).

\bibitem[{\citenamefont{David et~al.}(1998)\citenamefont{David, Reiter,
  Sitthai, and Schultz}}]{DRSS98}
\bibinfo{author}{\bibfnamefont{M.~O.} \bibnamefont{David}},
  \bibinfo{author}{\bibfnamefont{G.}~\bibnamefont{Reiter}},
  \bibinfo{author}{\bibfnamefont{T.}~\bibnamefont{Sitthai}}, \bibnamefont{and}
  \bibinfo{author}{\bibfnamefont{J.}~\bibnamefont{Schultz}},
  \bibinfo{journal}{Langmuir} \textbf{\bibinfo{volume}{14}},
  \bibinfo{pages}{5667} (\bibinfo{year}{1998}).

\bibitem[{\citenamefont{Renger et~al.}(2000)\citenamefont{Renger,
  M{\"u}ller-Buschbaum, Stamm, and Hinrichsen}}]{RMSH00}
\bibinfo{author}{\bibfnamefont{C.}~\bibnamefont{Renger}},
  \bibinfo{author}{\bibfnamefont{P.}~\bibnamefont{M{\"u}ller-Buschbaum}},
  \bibinfo{author}{\bibfnamefont{M.}~\bibnamefont{Stamm}}, \bibnamefont{and}
  \bibinfo{author}{\bibfnamefont{G.}~\bibnamefont{Hinrichsen}},
  \bibinfo{journal}{Macromolecules} \textbf{\bibinfo{volume}{33}},
  \bibinfo{pages}{8388} (\bibinfo{year}{2000}).

\bibitem[{\citenamefont{Morariu et~al.}(2003)\citenamefont{Morariu,
  Sch{\"a}ffer, and Steiner}}]{MSS03}
\bibinfo{author}{\bibfnamefont{M.~D.} \bibnamefont{Morariu}},
  \bibinfo{author}{\bibfnamefont{E.}~\bibnamefont{Sch{\"a}ffer}},
  \bibnamefont{and} \bibinfo{author}{\bibfnamefont{U.}~\bibnamefont{Steiner}},
  \bibinfo{journal}{Eur. Phys. J. E} \textbf{\bibinfo{volume}{12}},
  \bibinfo{pages}{375} (\bibinfo{year}{2003}).

\bibitem[{not()}]{note2}
 \bibinfo{note}{As pointed out to us by A. Sharma, a
  predecessor of Eqs.\,(\ref{TWOL_EV}) in terms of pressures rather than energy
  variations is given in Ref.\,\cite{Band01}. The system given in
  Ref.\,\cite{Dano98} can be transformed into the one of Ref.\,\cite{Band01}
  when neglecting surface viscosity.}

\bibitem[{\citenamefont{Brochard-Wyart
  et~al.}(1993)\citenamefont{Brochard-Wyart, Martin, and Redon}}]{BMR93}
\bibinfo{author}{\bibfnamefont{F.}~\bibnamefont{Brochard-Wyart}},
  \bibinfo{author}{\bibfnamefont{P.}~\bibnamefont{Martin}}, \bibnamefont{and}
  \bibinfo{author}{\bibfnamefont{C.}~\bibnamefont{Redon}},
  \bibinfo{journal}{Langmuir} \textbf{\bibinfo{volume}{9}},
  \bibinfo{pages}{3682} (\bibinfo{year}{1993}).

\bibitem[{\citenamefont{Reiter et~al.}(1996)\citenamefont{Reiter, Schultz,
  Auroy, and Auvray}}]{RSAA96}
\bibinfo{author}{\bibfnamefont{G.}~\bibnamefont{Reiter}},
  \bibinfo{author}{\bibfnamefont{J.}~\bibnamefont{Schultz}},
  \bibinfo{author}{\bibfnamefont{P.}~\bibnamefont{Auroy}}, \bibnamefont{and}
  \bibinfo{author}{\bibfnamefont{L.}~\bibnamefont{Auvray}},
  \bibinfo{journal}{Europhys. Lett.} \textbf{\bibinfo{volume}{33}},
  \bibinfo{pages}{29} (\bibinfo{year}{1996}).

\bibitem[{\citenamefont{Velev et~al.}(2003)\citenamefont{Velev, Prevo, and
  Bhatt}}]{VPB03}
\bibinfo{author}{\bibfnamefont{O.~D.} \bibnamefont{Velev}},
  \bibinfo{author}{\bibfnamefont{B.~G.} \bibnamefont{Prevo}}, \bibnamefont{and}
  \bibinfo{author}{\bibfnamefont{K.~H.} \bibnamefont{Bhatt}},
  \bibinfo{journal}{Science} \textbf{\bibinfo{volume}{426}},
  \bibinfo{pages}{515} (\bibinfo{year}{2003}).

\bibitem[{\citenamefont{Sharma}(1993)}]{Shar93}
\bibinfo{author}{\bibfnamefont{A.}~\bibnamefont{Sharma}},
  \bibinfo{journal}{Langmuir} \textbf{\bibinfo{volume}{9}},
  \bibinfo{pages}{861} (\bibinfo{year}{1993}).

\bibitem[{not()}]{note1}
 \bibinfo{note}{
The four-index Hamaker constants are calculated using
  an equivalent of Eq.\,(11.13) of Ref.\,\cite{Isra92} that is based on the
  assumption that the main absorption frequencies of all involved media are
  about $\nu = 3\times10^{15}$\,Hz and that the zero frequency contribution is
  neglegtable. This yields 
\begin{displaymath}
  A_{ijkl}\,\approx\,\frac{3h\nu_e}{8\sqrt{2}}\,
  \frac{(n_i^2-n_j^2)(n_l^2-n_k^2)}{(n_i^2+n_j^2)^{1/2}(n_l^2+n_k^2)^{1/2}%
  [(n_i^2+n_j^2)^{1/2}+(n_l^2+n_k^2)^{1/2}]} 
\end{displaymath}
The three-index Hamaker constants
  are obtained by $A_{ijk}=A_{ijjk}$. For the Si/PMMA/PS/air system one obtains
  $A_{12g} = 1.49 \times 10^{-20}$\,J, $A_{21s} = 3.81 \times 10^{-20}$\,J and
  $A_{g21s} = -23.02\times10^{-20}$\,J, whereas the SiO/PMMA/PS/air system is
  characterized by $A_{12g} = 1.49 \times 10^{-20}$\,J, $A_{21s} = -0.02\times
  10^{-20}$\,J and $A_{g21s} = 0.15\times 10^{-20}$\,J. The used refractive
  indices of the media are $n_{\mbox{PS}} = 1.59$, 
$n_{\mbox{PDMS}} = 1.43$, $n_{\mbox{PMMA}} = 1.49$,
  $n_{\mbox{Si}} = 4.11$ and $n_{\mbox{SiO}} = 1.48$ \cite{MSS03}. 
We generally neglect acoustic
  contributions to the disjoining pressure \cite{ScSt02} because the accoustic
  modes are not confined to the individual liquid layers \cite{MSS03}.}

\bibitem[{\citenamefont{Bandyopadhyay}(2001)}]{Band01}
\bibinfo{author}{\bibfnamefont{D.}~\bibnamefont{Bandyopadhyay}},
  \emph{\bibinfo{title}{Stability and dynamics of bilayers}}, 
  \bibinfo{note}{Master-thesis, Ind. Inst. Tech. Kanpur}, (\bibinfo{year}{2001}). 

\bibitem[{\citenamefont{Danov et~al.}(1998)\citenamefont{Danov, Paunov,
  Alleborn, Raszillier, and Durst}}]{Dano98}
\bibinfo{author}{\bibfnamefont{K.~D.} \bibnamefont{Danov et~al.}},
  \bibinfo{journal}{Chem. Eng. Sci.} \textbf{\bibinfo{volume}{53}},
  \bibinfo{pages}{2809} (\bibinfo{year}{1998}).



\bibitem[{\citenamefont{Sch{\"a}ffer and Steiner}(2002)}]{ScSt02}
\bibinfo{author}{\bibfnamefont{E.}~\bibnamefont{Sch{\"a}ffer}}
  \bibnamefont{and} \bibinfo{author}{\bibfnamefont{U.}~\bibnamefont{Steiner}},
  \bibinfo{journal}{Eur. Phys. J. E} \textbf{\bibinfo{volume}{8}},
  \bibinfo{pages}{347} (\bibinfo{year}{2002}).

\bibitem[{\citenamefont{Thiele, John and B{\"a}r}(2004)}]{ThJB04}
\bibinfo{author}{\bibfnamefont{U.}~\bibnamefont{Thiele}}
  \bibinfo{author}{\bibfnamefont{K.}~\bibnamefont{John}}
  \bibnamefont{and} \bibinfo{author}{\bibfnamefont{M.}~\bibnamefont{B{\"a}r}},
  \bibinfo{journal}{Phys. Rev. Lett.} \textbf{\bibinfo{volume}{93}},
  \bibinfo{pages}{027802} (\bibinfo{year}{2004}).
\end{thebibliography}

%
\begin{figure}[H]
\caption{Geometry of the two-layer system. The mean film thicknesses of the
lower and upper layer are $d_1$ and $d_2-d_1$, respectively.} 
\label{fig1}
\end{figure}
\begin{figure}[H]
\caption{Snapshots from time evolutions of a two-layer film for a
Si/PMMA/PS/air system at dimensionless times (in units of $\tau_{\text{up}}$) 
as shown in the insets.
(a) At $d= 1.4$ a varicose mode evolves leading to rupture
of the upper layer at the liquid-liquid interface. The ratio of the time 
scales derived from upper and lower effective one-layer system is
$\tau_{\text{up}}/\tau_{\text{low}} = 0.066$.
(b) At 
$d = 2.4$ a zigzag mode evolves and rupture of the lower layer 
occurs at the substrate ($\tau_{\text{up}}/\tau_{\text{low}} = 34.98$).
The domain lengths are 5 times the corresponding fastest unstable 
wave length and $\mu=\sigma=1$. 
}
\label{fig4}
\end{figure}
\begin{figure}[H]
\caption{Stability diagram for fixed scaled coupling 
$\partial_{h_1h_2}\rho_{VW}= 8 \pi^2 A_{12g}/|A_{12g}|$.
Shown are the stability threshold 
(solid line) and the boundary
between unstable one-mode and two-mode regions (dashed line). 
The thin lines represent the trajectories for commonly studied systems: 
(1) Si/PMMA/PS/air, (2) SiO/PMMA/PS/air, (3) SiO/PS/PDMS/air, (4)
Si/PS/PDMS/air, and (5) Si/PDMS/PS/air.  The Hamaker 
constants were calculated as detailed below \cite{note1}. 
}
\label{fig2}
\end{figure}
\nopagebreak
\begin{figure}[H]
\caption{Dispersion relations $\beta(k)$ 
for a Si/PMMA/PS/air system at $d=1.4$ (dashed line, varicose mode) and $d=2.4$
(dot-dashed line, zigzag mode) for parameters as in Fig.\,\ref{fig4}; and
$4\beta(k)$ for a SiO/PMMA/PS/air system at $d = 2.16$ (solid line)
for $\mu=1$ and $\sigma=10$.
}
\label{fig3}
\end{figure}
\begin{figure}[H]
\caption{Single snapshots from time evolutions of a 
SiO/PMMA/PS/air system for $d = 2.16, \mu=1$ 
and different $\sigma$.
(a) The respective evolutions of the two interfaces are dominated 
by modes of different wavelength 
($\sigma=10, t=3.83$). 
In (b) and (c) the evolution 
is dominated by the liquid-gas and the liquid-liquid interface, 
respectively ($\sigma=5, t=3.9$ and $\sigma=100, t=0.52$).
}
\label{fig5}
\end{figure}


%
\clearpage
\begin{center}
\includegraphics[width=0.8\hsize]{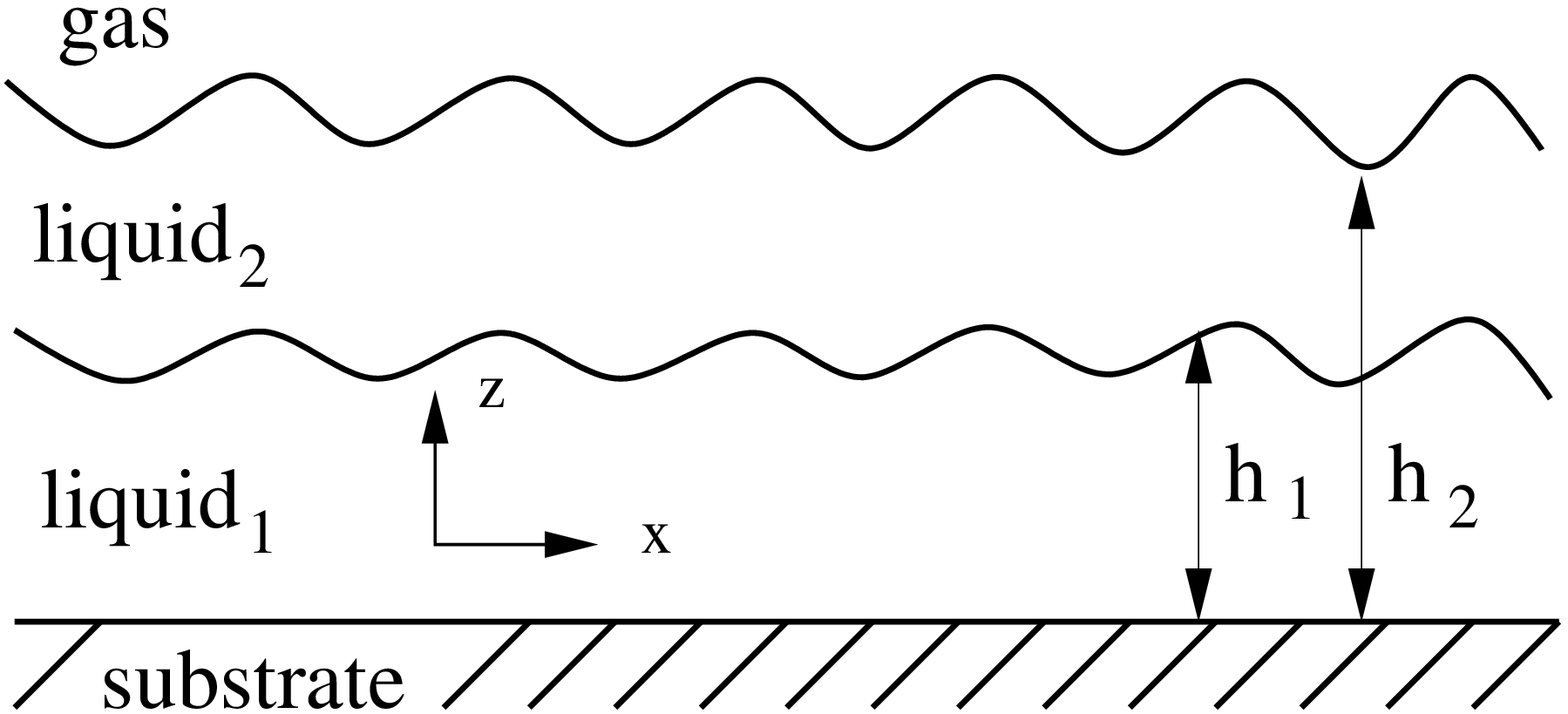}                           
\vspace{4cm}

{\bf\large Fig.\,\ref{fig1}\\
Pototsky et al., PRE}
\end{center}

\newpage

\begin{center}
\includegraphics[width=0.9\hsize]{fig2_mod.eps} 
\vspace{4cm}

{\bf\large Fig.\,\ref{fig4}\\
Pototsky et al., PRE}
\end{center}

\newpage

\begin{center}
\includegraphics[width=0.9\hsize]{phase2_mod.eps}
\vspace{4cm}

{\bf\large Fig.\,\ref{fig2}\\
Pototsky et al., PRE}
\end{center}

\newpage

\begin{center}
\includegraphics[width=0.7\hsize]{disp_betas.eps}
\vspace{4cm}

{\bf\large Fig.\,\ref{fig3}\\
Pototsky et al., PRE}
\end{center}

\newpage

\begin{center}
\includegraphics[width=0.8\hsize]{fig5_mod.eps}
\vspace{4cm}

{\bf\large Fig.\,\ref{fig5}\\
Pototsky et al., PRE}
\end{center}

\end{document}